\documentstyle[sprocl,phase1]{article}
\input{psfig}
\def\Journal#1#2#3#4{{#1} {\bf #2}, #3 (#4)}

\def\PLB{{\em Phys. Lett.}  B}

\def\PRD{{\em Phys. Rev.} D}

\def\mco{\multicolumn}

\def\ra{\rightarrow}

\def\ko{K^0}

\def\be{\begin{equation}}
\def\ee{\end{equation}}
\def\bea{\begin{eqnarray}}
\def\eea{\end{eqnarray}}

\def\deg{\hbox{$^\circ$}}
\begin{document}

\

\vskip3cm

\Large
\centerline{\bf ~~~ Measuring Cosmological Parameters }

\bigskip
\bigskip
\bigskip
\bigskip

\centerline{~~~Wendy L. Freedman$^1$}
\normalsize 

\bigskip
\vskip3cm
\vskip3cm

\noindent
$^1$ Carnegie Observatories, 813 Santa Barbara St., Pasadena, CA 91101.

\bigskip
\bigskip

\noindent
{\it Invited Review given at the 18th Texas Symposium held in Chicago,
December 1996. To be published by World Scientific Press, eds. A.
Olinto, J. Frieman, and D. Schramm.}

\smallskip
\noindent
February, 1997.

\vfill\eject

\title{MEASURING COSMOLOGICAL PARAMETERS}
\author{ Wendy L. FREEDMAN}
\address{Carnegie Observatories, 813 Santa Barbara St., Pasadena, \\CA
91101, USA}
\maketitle\abstracts{
In  this  review,  the status  of  measurements  of the matter density
($\Omega_m$),  the   vacuum  energy density  or  cosmological constant
($\Omega_\Lambda$), the   Hubble constant (H$_0$),  and   ages  of the
oldest measured objects  (t$_0$) are summarized.  Measurements  of the
statistics  of gravitational  lenses  and strong gravitational lensing
are discussed in the  context  of limits on $\Omega_\Lambda$.    Three
separate routes to the Hubble constant are considered: the measurement
of  time  delays in  multiply-imaged   quasars, the Sunyaev-Zel'dovich
effect  in   clusters,    and  Cepheid-based extragalactic  distances.
Globular-cluster ages plus a new age measurement  based on radioactive
dating of thorium in a metal-poor star are briefly summarized.  Limits
on the product   of $\rm H_0t_0$  are  also  discussed.  Many  recent,
independent dynamical measurements are  yielding a low  value  for the
matter density ($\rm  \Omega_m \sim$  0.2-0.3). A wide range of Hubble
constant measurements appear to  be converging in the  range  of 60-80
km/sec/Mpc.   Areas where future improvements  are  likely to be  made
soon are  highlighted; in particular,  measurements of anisotropies in
the cosmic microwave   background.  Particular   attention is paid  to
sources of systematic error and  the assumptions that underlie many of
the measurement methods.}


\section{Introduction }

Rapid progress is being made in measuring the cosmological parameters
that describe the dynamical evolution and the geometry of the Universe.
In essence,  this is the first  conclusion of this review.  The second
conclusion is that despite the  considerable advances, the accuracy of
cosmological parameters is not  yet sufficiently high to  discriminate
amongst, or  to  rule out  with confidence, many existing, competing, world
models.  We as observers still  need to do better.  Fortunately, there
are a number of opportunities on the horizon that will  allow us to do
so.


In the context of the general theory of relativity, and assumptions of 
large-scale homogeneity and isotropy, the dynamical evolution of the Universe
is specified by the Friedmann equation

$$ \rm H^2  = { 8\pi G \rho_m \over 3} - { k \over a^2}  +  { \Lambda \over 3}$$

\noindent
where a(t) is  the scale factor, H=${\dot{a} \over a}$  is the Hubble
parameter (and H$_0$ is the Hubble ``constant'' at the present epoch),
$\rho_m$  is  the average  mass density, k  is  a curvature term,  and
$\Lambda$ is the cosmological  constant, a  term which  represents the
energy density of  the vacuum.  It  is  common practice to  define the
matter density ($\rm  \Omega_m  = 8\Pi  G\rho_m /  3H_0^2)$, the vacuum
energy density ($\rm  \Omega_\Lambda  = \Lambda  / 3H_0^2$),  and the
curvature term ($\rm \Omega_k$ = -k /a$_0^2$H$_0^2$) so that $\rm
\Omega_m$ + $\rm \Omega_\Lambda$  = 1 for the case  of a flat universe
where k = 0.  The simplest case  is  the Einstein-de Sitter model with
$\rm  \Omega_m$ = 1 and $\rm  \Omega_\Lambda$ = 0.   The dimensionless
product $\rm H_0t_0$ (where $\rm t_0$ is the age of the Universe) is a
function of both  $\rm \Omega_m$ and $\rm  \Omega_\Lambda$. In the case
of the Einstein-de Sitter Universe

$$ \rm f(\Omega_m, \Omega_\Lambda) =  H_0t_0 =  {2 \over 3} $$

\begin{figure} 
\psfig{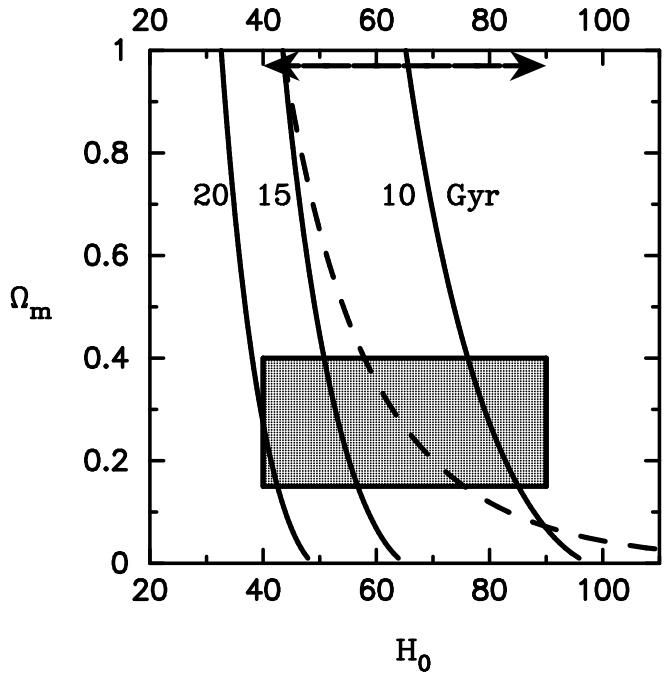} \caption{
$\rm  \Omega_m$ versus $\rm  H_0$ showing
current observational limits on cosmological parameters.
Solid lines denote expansion ages for  an open ($\rm
\Omega_\Lambda$ = 0) Universe and the dashed line denotes
an expansion age of 15 Gyr in the case of a  flat ($\rm \Omega_\Lambda
\neq$ 0) Universe.  See text for details.  }
\end{figure}

Bounds on  several cosmological parameters  are summarized in Figure 1
in a plot of the matter density as a function of the  Hubble constant,
following Carroll, Press \& Turner (1992).  Solid lines  represent the
expansion ages for 10, 15, and 20 Gyr in  an open ($\rm  \Lambda$ = 0)
model.  The grey box is defined by values of H$_0$ in  the range of 40
to 90 km/sec/Mpc  and  0.15 $\rm <  \Omega_m <$  0.4. The solid  arrow
denotes the same range in $\rm H_0$ for $\rm
\Omega_m$ = 1. This plot  illustrates  the well-known ``age'' problem;
namely that for an Einstein-de Sitter Universe ($\rm
\Omega$ =  1, $\rm \Lambda$ = 0),  H$_0$ must  be less  than $\sim$45
km/sec/Mpc   if  the ages  of  globular   clusters  (t$_0$) are indeed
$\sim$15 billion  years old.  This  discrepancy  is less severe if the
matter  density of the Universe is  less than the critical density, or
if  a non-zero  value of  the  cosmological constant  is allowed.  For
example, the dashed line indicates an expansion age of  15  Gyr in the
case of  a flat ($\rm  \Omega_m +  \Omega_\Lambda$ = 1)  model for $\rm
\Lambda \neq 0$. 

A  number  of issues  that   require   knowledge of  the  cosmological
parameters remain unresolved  at present.   First is  the  question of
timescales ($\rm H_0t_0$) discussed above; possibly a related issue is
the observation  of  red (if they  are  indeed old) galaxies  at  high
redshift.  Second is  the amount of  dark matter in the Universe.   As
discussed  below, many dynamical  estimates  of the mass  over  a wide
range of scale sizes are currently favoring values of $\Omega_m
\sim$0.25$\pm$0.10,  lower  than the  critical  Einstein-de  Sitter
density.  And third  is the  origin  of large-scale structure  in  the
Universe.   Accounting  for  the  observed  power spectrum  of  galaxy
clustering  has turned  out  to be  a   challenge to the best  current
structure formation models.

Taking the current data at face value, there appears to  be a conflict
with the standard Einstein-de Sitter model.  In  fact, it is precisely
the resolution of these problems that has  led  to a recent resurgence
of interest in a non-zero value of  $\Lambda$ ({\it e.g.}  Ostriker \&
Steinhardt 1995; Krauss \& Turner 1995).  Another means  of addressing
these issues ({\it e.g.} Bartlett {\it et al.} 1995) requires being in
conflict   with   essentially   all  of   the   current  observational
measurements of H$_0$; from purely theoretical considerations,  a very
low value of H$_0$ ($\leq$30) could also resolve these issues.

Ultimately we will have to defer to measurement as the arbiter amongst
the   wide range  of  cosmological   models (and their  very different
implications) still being discussed in the literature. A wealth of new
data is   becoming available  and progress  is   being   made  in  the
measurement of all of the cosmological parameters discussed below: the
matter  density,  $\rm \Omega_m$,   the vacuum  energy   density, $\rm
\Omega_\Lambda$, the expansion rate $\rm  H_0$, and age  of the oldest
stars  $\rm t_0$.  The central, critical  issues  now are (and in fact
have  always been) testing  for and  eliminating sources of significant
systematic error.

\section{ $\bf \Omega_m$ -- The Matter Density}

Table 1   presents  a  summary  of several  different   techniques for
measuring the matter  density of the Universe.  These  techniques have
been developed over a wide range of scales, from galaxy ($\sim$100-200
kpc), through cluster (Mpc), on up to more global scales (redshifts of
a few).  Excellent,  recent reviews on  determinations of $\rm \Omega$
can  be found in  Dekel, Burstein \&  White (1997) and Bahcall, Lubin
\& Dorman 1995 and references therein.  The first part of
the table lists $\rm \Omega_m$ determinations that are  independent of
$\rm  \Omega_\Lambda$;   the second   part   lists    $\rm   \Omega_m$
determinations that are not independent  of $\rm  \Omega_\Lambda$; and
the third part of the table lists $\Omega_\Lambda$ determinations.  In
addition to listing the physical basis of  the method, types of object
under study, and values  of $\Omega_m$ plus an estimated  uncertainty,
Table 1 makes explicit some  of the assumptions  that underlie each of
these techniques.  Although in many cases, 95\%  confidence limits are
quoted, these estimates must ultimately be evaluated in the context of
the validity  of their  underlying assumptions.   It is non-trivial to
assign  a  quantitative   uncertainty  in many  cases,   but   in fact
systematic effects may be the dominant source of uncertainty.  Several
of  these assumptions and  uncertainties  are discussed further below.
They  include, for example,   diverse  assumptions about  mass tracing
light,   mass-to-light  ratios     being    constant,  clusters  being
representative of the Universe, clumping  of  X-ray gas, non-evolution
of type Ia supernovae,  and the  non-evolution of elliptical galaxies.
For methods that operate over very large scales (gravitational lensing
and type Ia  supernovae), assumptions about $\rm \Omega_{\Lambda}$  or
$\rm
\Omega_{total}$ are currently required to place limits on $\rm
\Omega_m$.

\begin{table}
\vspace{-1cm}
  \caption{SUMMARY OF $\Omega_m$ and $\Omega_\Lambda$ DETERMINATIONS }
\medskip
\medskip
\normalsize
\begin{tabular}{|l|l|l|l|l|l|} 
\hline
 Sample & Method & Scale & Assumptions & $\LARGE \bf \Omega_m$ & Error   \\[3pt] 
\hline \mco{5}{|c|}{$ \bf \Omega_\Lambda$ Independent Methods}  &\\ 
\hline
 Galaxies & dyn. M/L ratio   &  100 kpc & galaxies representative & $\sim$0.1  &   \\
                               &    &   & M/L constant  &  &   \\
 Clusters & dyn. M/L ratio  &  $<$few Mpc & clusters representative & $\sim$0.2  &   \\
                                  &   &  & M/L constant &  &   \\
 Clusters & X-ray M/L ratio   &  $<$few Mpc & hydrostatic eqm & $\sim$0.2  &   \\
 Clusters   & baryon fraction &  & clusters representative  &  0.3-0.5  &   \\
                               &  &  & no clumping  &  &   \\
 Clusters   & morphology &  & model dept. &  $>$0.3  &   \\
 Local Group   & Least action principle & 1 Mpc & LG representative  & $\sim$0.15   &   \\
                                          &  &  & no external torques  &    &   \\
                                          &  &  & model uniqueness  &    &   \\
 Galaxies  &  Virial Theorem  & 1-300 Mpc &  mass-indept. biasing  & 0.2-0.4  &   \\
                                           &   (pairwise velocities)  &  & point masses &  &   \\
 Galaxies    & Peculiar velocities & 100 Mpc & biasing  & $>$0.3   & 95\%  \\
\hline
    &    &   &    &  &  \\

    &    &   &    &  & \\
  Sample & Method & Scale & Assumptions & $\LARGE \bf \Omega_m $ & Error   \\[3pt]
\hline \mco{5}{|c|}{$ \bf \Omega_\Lambda$ Dependent Methods}   &\\ 
\hline
 Type Ia SNae   & Hubble diagram   & z$<$0.5  &  $\Omega_\Lambda $= 0  &  0.88  & 90\% \\
    &    &   &  no evolution effects  &   &  \\
    &    &   &  $\Omega_{tot}$=1  & $>$0.49   & 95\% \\
 Lensed QSO's   & lensing statistics   &  global & $\Omega_{\Lambda}$=0 & $>$0.15 & 90\%\\
                                   &    &   & dark matter distrib. &   &  \\
                                   &    &   & slow galaxy evolution  &  &  \\
                                   &    &   & dust small effect    &  & \\
 6 lenses   & strong lensing   & global  & $\Omega_\Lambda$ = 0   & low $\Omega$  &  \\
   &    &   &  model dependent  &   &  \\

 CMB    &  multipole analysis & global  & CDM   & 0.3 - 1.5    &   \\
\hline
    &    &   &    &  &  \\
    &    &   &    &  &  \\
 Sample & Method & Scale & Assumptions & $\LARGE \bf \Omega_\Lambda$ & Error   \\[3pt]
\hline
 Type Ia SNae   &    & z$<$0.5  &  $\Omega_{tot}$=1  &  $<$0.51  & 95\% \\
 Lensed QSO's   & lensing statistics   &  global &   $\Omega_{tot}$=1 & $<$ 0.66 & 95\% \\
 6 lenses   & strong lensing   & global  &  $\Omega_{tot}$=1  &  $<$0.9 & 95\% \\
 H$_0$t$_0$   & age discrepancy   & 100 Mpc  & H$_0 > $65   & $>$0.5 & 66\% \\
   &   &   & t$_0 >$ 13 Gyr    &  &  \\
\hline

\end{tabular}

\end{table}








Since lower  values of   the matter density  tend  to be  measured  on
smaller  spatial scales, it has  given rise to the suspicion  that the
true, global value of $\rm \Omega_0$ must be measured on scales beyond
even  those of large clusters,   {\it  i.e.,}  scales of greater  than
$\sim$100 Mpc ({\it   e.g.}, Dekel 1994).    In that  way,   one might
reconcile the low  values of  $\rm \Omega_m$  inferred locally  with a
spatially flat Universe.   However, recent  studies (Bahcall, Lubin \&
Dorman  1995) suggest that the M/L  ratios of galaxies do not continue
to grow beyond a scale size of  about $\sim$200  kpc (corresponding to
the sizes of  large halos   of individual galaxies).    In their Jeans
analysis of the dynamics of  16  rich clusters, Carlberg {\it  et al.}
(1997) also  see no further  trend with scale.   Hence, currently  the
observational evidence does not indicate that measurements of $\rm
\Omega_m$ on cluster  size scales are biased to  lower values than the
true global value.

A brief description of several  techniques  for measuring  the  matter
density is given below. These methods are discussed in the  context of
both  their  strengths and weaknesses, paying  particular attention to
the underlying assumptions.  An excellent and more  complete review on
this topic is   given by Dekel,   Burstein \& White  (1997); also  see
Trimble (1987).

\subsection{ Galaxies and Clusters: Dynamical Measures \& Mass-to-Light Ratios }

The contribution of galaxies to the mass density  can be determined by
integrating the luminosity function per unit volume  for  galaxies and
multiplying by an (assumed, constant)  mean mass-to-light (M/L) ratio.
The dynamical masses  of   galaxies can be determined   from  rotation
curves for spiral galaxies, or the measurement of velocity dispersions
and application of the  virial  theorem both for individual elliptical
galaxies.  The  latter   method can  also  be applied  for groups  and
clusters of galaxies (as Zwicky did in the 1930's).

This method has several  advantages.   First it is conceptually simple
and model-independent.  Unlike some of the global techniques discussed
below, this method is independent of both $\rm H_0$ and $\rm
\Omega_\Lambda$.    However,    there are  a    number  of  underlying
assumptions.   Most  important  is {\it  the  assumption that galaxies
trace    all mass}.   In addition,   there  are   implicit, underlying
assumptions  concerning  the similarity of    mass-to-light ratios  in
different  systems (ignoring,  for  example,  potential differences in
initial   mass functions,  star  formation histories,    dark  remnant
populations, dust content,  etc.) The estimates  based on  this method
tend to yield low values of $\rm \Omega_m$ of $\leq$ 0.25.

\subsection{ Dynamics of the Local Group }

Peebles  (1994)    estimated $\rm
\Omega_m$ by calculating the orbits of galaxies in the Local Group
based on observed radial velocities,  positions, and  distances. Shaya
{\it et al.} (1995)  extended this  method to  a catalog   of galaxies
within 3000  km/sec.  Again  this  is a   method that  is conceptually
straightforward and independent of $\rm H_0$ and $\rm \Omega_\Lambda$.
Moreover, since the galaxies are nearby, the  errors  in the distances
are relatively small.  However, only one (the radial) component of the
motion is measured.  This  method too is based  on {\it the assumption
that galaxies  trace  mass}.   It  also assumes that  external   tidal
influences  and past mergers  are not significant.   Furthermore,  the
question  of uniqueness is  difficult to address.  The estimates based
on this method again give low values of $\rm \Omega_m$ of $\sim$ 0.15.

\subsection{ Cluster Baryon Fraction }

This issue was discussed  in detail by White  {\it  et al.} (1993) for
the Coma cluster, and has been addressed now in many contexts by a
number of  authors ({\it e.g.}, White \&  Frenk 1991; White  \& Fabian
1995;  Steigman \& Felten   1995).  The calculation goes  as  follows:
First,   the  number density  of  baryons  ($\rm  \Omega_b$)  can   be
determined based  on   the observed densities   of light elements from
big-bang nucleosynthesis.  Hence, the fraction of baryons $ \rm (f_b)$
measured  in clusters  of  galaxies can be used    to estimate of  the
overall matter density assuming

$$ \rm f_b = { M_{gas} \over M_{TOT} } = { M_b \over M_{TOT} } =  { \Omega_b \over \Omega_{m} } $$
 
\par\noindent

There are four explicit assumptions made:

1) The gas is in hydrostatic equilibrium.

2) There is a smooth potential.

3) Most of the baryons in the clusters are in the X-ray gas.

4) The cluster baryon fraction is representative of the Universe.

\medskip


\noindent
If the gas is clumped or there is another source of pressure (magnetic
fields or turbulence) in addition to the  thermal pressure, the baryon
fraction would be decreased and the matter  density would be increased
(Steigman \& Felten 1995).

Recent  measurements of  X-ray clusters  ({\it  e.g.}, Loewenstein  \&
Mushotsky   1996;  White \&  Fabian   1995) indicate  that the  baryon
fraction  has a range  of values from  about  10-$>$20\%.  The values for
$\rm f_b$ tend to be smaller for small groups and in the inner regions
of larger  clusters.   These  results  underscore the  importance   of
ensuring that such measurements are made on large  enough scales to be
truly representative of the large-scale Universe as a whole.

Taken at face  value, the cluster-baryon  method estimates again favor
low values of $\rm \Omega_m$. For $\rm \Omega_bh^2$ = 0.024 $\pm$ 12\%
(Tytler, this conference) relatively low  values  of $\rm \Omega_m < $
0.5 are favored for  the  range  of  baryon  fractions observed.   The
Tytler {\it  et al.} 1997 baryon determination  is at the high end  of
recent measures of this quantity (low  end of  the deuterium abundance
measurements); lower baryon densities only serve  to decrease the $\rm
\Omega_m$  estimates.  (However, see the discussion   by Bothun, Impey
and     McGaugh    1997;    these   authors    suggest  that   perhaps
low-surface-brightness galaxies could be source of most of the baryons
in the Universe  and that rich clusters  are not representative of the
overall baryon density.)


\subsection{ Peculiar Velocities:  Density and Velocity Comparisons  }

On scales of $\sim$100 Mpc, the motions of field galaxies can  be used
to infer  the mass   density given  independent  distance information.
These methods do not yield a  measure  of $\rm \Omega_m$ directly, but
rather yield the ratio $\rm \beta = \Omega^{0.6}  / b$  where b is the
bias parameter (describing the relation between mass and light) over a
scale of a few hundred km/sec.  These methods are again insensitive to
both  H$_0$ and  $\Omega_\Lambda$.  Several different  approaches have
been investigated.  For more details,  the reader is referred to Dekel
(1994), Willick {\it  et  al.} (1997)  and Dekel, Burstein   and White
(1997).


All methods make use of  radial  velocity catalogs and distances based
on the Tully-Fisher relation. The analyses differ  in detail and there
are  advantages and disadvantages  to each type  of approach.  At  the
present time, the results from  this type  of technique  have not  yet
yielded a consistent picture.  Earlier analyses ({\it e.g.} Dekel {\it
et al.} 1993)  suggested large values of  $ \rm \beta   \sim$ 1.3, and
correspondingly   rather high values   of   $  \rm \Omega$ (subject to
assumptions about the value of b).   More  recently,  the estimates of
$\rm \beta$  have decreased somewhat  (Dekel, Burstein \& White 1997).
At present, the  results  from different  groups ({\it  e.g.},  Dekel,
Willick, Davis and collaborators) appear to differ from the results of
Giovanelli, Haynes, Da  Costa and  collaborators (see the contribution
by Da  Costa   to  this volume).   Understanding   the sources of  the
differences is clearly an important goal.

\subsection{ Galaxy Pairwise Velocities }

Using the cosmic virial  theorem, the relative  velocity dispersion of
galaxy  pairs can be used  to estimate the  matter density ({\it e.g.}
Davis  \&  Peebles 1983).  The  Las Campanas Redshift Survey (Shectman
{\it et al.} 1996) contains about 26,000 redshifts out to $\sim$30,000
km/sec and provides an excellent sample of  galaxies not  dominated by
clusters.  Davis   (this conference) presented   results based on this
sample, concluding that relative  galaxy pairs have  a one-dimensional
velocity dispersion of  only 260 km/sec,  implying $\rm \Omega_m \sim$
0.25.

This method  is very clean  and conceptually simple; however, it again
is limited   by the assumption  that bias   is  independent of  scale.
Moreover,  Frenk   (1997)  argues that   bulk velocity flows   are not
sensitive  to $\rm  \Omega_m$, and  that  the peculiar  velocities are
quite similar for a number of models with a range of values of $\rm
\Omega_m$.

\section{ $\bf \Omega_\Lambda$ and $\bf \Omega_m$ Limits}

The subject of the cosmological constant $\Lambda$  has had a long and
checkered history in cosmology. The reasons for skepticism regarding a
non-zero value of the cosmological constant are many.  First, there is
a  discrepancy of   $\geq$120  orders  of  magnitude  between  current
observational limits and estimates  of the vacuum energy density based
on current  standard  particle theory ({\it  e.g.}  Carroll, Press and
Turner 1992).  Second,  it would require  that  we are now living at a
special epoch when  the cosmological constant has begun  to affect the
dynamics of the Universe (other than during a  time of inflation).  In
addition,  it   is difficult  to ignore  the fact  that historically a
non-zero  $\Lambda$  has been  dragged out prematurely  many  times to
explain a  number  of  other apparent   crises, and  moreover,  adding
additional free parameters to a problem  always makes it easier to fit
data.  Certainly the oft-repeated  quote from Einstein to  Gamov about
his  ``biggest  blunder" continues  to undermine  the credibility of a
non-zero value for $\Lambda$.

However, despite the strong arguments that can be made for $\Lambda$ =
0, there  are compelling  reasons to keep an  open  mind on the issue.
First,  at present there is no  known physical principle that  demands
$\Lambda$ = 0.  Although supersymmetry can provide  a mechanism, it is
known that  supersymmetry  is  broken   ({\it  e.g.,} Weinberg  1989).
Second,  unlike the case   of Einstein's  original  arbitrary constant
term, standard particle  theory and  inflation now provide a  physical
interpretation of $\Lambda$: it is  the  energy density of  the vacuum
({\it e.g.},    Weinberg   1989).  Third,   {\it  if}  theory  demands
$\Omega_{total}$= 1,  then a  number  of observational results  can be
explained with a low $\Omega_m$ and $\Omega_m +
\Omega_\Lambda  =  1$: a)  for instance, the  observed  large  scale distribution of galaxies, clusters,  large voids,  and walls is in conflict  with that
predicted by the (standard) cold dark  matter  model for the origin of
structure ({\it e.g.} Davis {\it et al.} 1992; Peacock \& Dodds 1994);
and b) the low values  of  the  matter density  based  on a number  of
methods as described in  $\S$2.  In addition, the discrepancy  between
the ages of  the oldest stars  and the expansion  age can be resolved.
Perhaps the most important reason to keep an open mind is that this is
an issue that ultimately must be resolved by experiment.

The importance of empirically establishing whether there is a non-zero
value of $\Lambda$ cannot be overemphasized.   However, it underscores
the need for high-accuracy experiments: aspects  of the standard model
of particle theory have  been tested in  the laboratory to  precisions
unheard of  in    most  measurements  in    observational   cosmology.
Nevertheless,  cosmology  offers an opportunity   to test the standard
model over larger scales and higher energies than can ever be achieved
by  other means.  It scarcely needs  to be said that  overthrowing the
Standard  Model ( {i.e.,} claiming  a measurement of  a non-zero value
for $\Lambda$ )  will  require  considerably higher accuracy  than  is
currently available.

What are the current observational limits on $\Omega_\Lambda$?
In the next sections, limits based on both the observed numbers of
quasars multiply imaged by galaxy ``lenses''  and limits from a
sample of strongly lensed galaxies are briefly discussed.

\subsection{ Gravitational Lens Statistics}

Fukugita, Futamase \& Kasai (1990)  and Turner (1990) suggested that a
statistical study of the number density  of gravitational lenses could
provide a powerful test of a non-zero $\Lambda$. Subsequently a number
of studies  have been undertaken ({\it e.g.}  Fukugita \& Turner 1991;
Bahcall  {\it  et al.}   1992; Maoz {\it et  al.} 1993; Kochanek  1993,
1996).  The basic idea  behind this  method is  simple: the  number of
gravitationally  lensed objects   is   a very  sensitive function   of
$\Omega_\Lambda$.  For larger  values of $\Omega_\Lambda$,  there is a
greater probability that  a quasar  will be lensed because  the volume
over a given redshift interval is increased.  In  a flat universe with
a value of $\Omega_\Lambda$  = 1, approximately  an order of magnitude
more gravitational   lenses  are predicted  than  in a   universe with
$\Omega_\Lambda$ = 0 (Turner 1990).  Thus, simply counting the numbers
of gravitationally  lensed quasars can  provide a very powerful limit on
the  value  of $\Omega_\Lambda$.   In  practice, however,  there are a
number of complications:   galaxies  evolve (and perhaps  merge)  with
time,  even elliptical  galaxies contain dust,   the properties of the
lensing galaxies are not  well-known  (in particular, the  dark matter
velocity  dispersion is unknown), and  the  numbers of lensing systems
known at   present is very  small ($\sim$  20).   Moreover,  while the
predicted effects are very large for $\Omega_\Lambda$ = 1, because the
numbers are such a sensitive function of $\Omega_\Lambda$, it  is very
difficult to provide  limits below a  value  of about 0.6, given these
complicating effects.

Kochanek (1996) has recently discussed  these various effects in  some
detail, and investigated the  sensitivity of the  results to different
lens models and extinction.   His best estimated  limits to date are :
$\Omega_\Lambda   <$ 0.66   (95\%   confidence)   for  $\Omega_m$    +
$\Omega_\Lambda$ =   1, and $\Omega_m$   =  0.15 (90\%  confidence) if
$\Omega_\Lambda$ = 0.  Significant improvements to these  limits could
be made by increasing the size of the current lens samples.

\subsection{ Strong Gravitational Lenses }

A number of strong (elliptical galaxy) gravitational  lens systems are
known  that may  offer  the potential of  constraining   the  value of
$\Omega_m$ and     $\Omega_\Lambda$  through modeling    of  the  lens
properties.   This method is less sensitive  to $\Omega_\Lambda$ than
the statistics  of lensing, and again it  is sensitive to a  number of
possible systematic    effects: possible  perturbations     by cluster
potentials, uncertainties in  the underlying properties of the lensing
galaxies,   and  model-dependent corrections  due  to evolution.   The
objects are  faint and the errors  in  the luminosities   and velocity
dispersions are potentially very significant.  A recent analysis  of 7
strong  lenses has been  undertaken by Im {\it et   al.} (1996). Their
current results yield $\Omega_\Lambda $ = 0.64$^{+0.15}_{-0.26}$ ({\it
i.e.,}  this measurement  sits  almost  at the end  of the  range {\it
excluded} by  Kochanek (1996) at 95\%  confidence.   Im {\it   et al.}
exclude $\Omega_m$  =   1.0 at   97\%   confidence.   


\subsection{ $\rm \Omega_m$ and $\rm \Omega_\Lambda$ from Type Ia Supernovae}

The use of type Ia supernovae for measuring cosmological parameters is
covered elsewhere in this volume by Filippenko (nearby  supernovae and
determinations of H$_0$)  and  by Perlmutter  (distant supernovae  and
$\rm \Omega_m$ and $\rm \Omega_\Lambda$).  Hence,  these  objects will
not  be discussed  in much  detail  here,  except to highlight   their
potential,  and to summarize  some of the main difficulties associated
with them so that they  can be compared relative  to some of the other
methods discussed in this review.

The obvious advantage of type Ia supernovae is the small dispersion in
the Hubble  diagram, particularly after  accounting for differences in
the overall shapes or slopes of the light curves (Phillips 1993; Hamuy
{\it  et al.} 1995: Reiss,  Press  \&  Kirshner 1997).  In principle,
separation of the effects  of  deceleration  or a  potential  non-zero
cosmological constant is  straightforward,  provided that (eventually)
supernovae at redshifts of order unity can be measured with sufficient
signal-to-noise  and resolution against the  background  of the parent
galaxies.   The differences in  the observed effects of $\rm \Omega_m$
and  $\rm \Omega_\Lambda$  become increasingly  easier to  measure  at
redshifts exceeding $\sim$0.5.  In principle, the evolution  of single
stars should be simpler than  that of entire  galaxies (that have been
used for such measurements in the past).

At  the  present   time,  however,   it  is  difficult to  place   any
quantitative limits on the expected  evolutionary effects for type  Ia
supernovae since the  progenitors for these  objects have not yet been
unequivocally   identified.    Moreover,   there    may be   potential
differences  in the chemical  compositions of  supernovae observed now
and those observed at  earlier epochs.  In principle, such differences
could  be tested  for  empirically  (as  is  being  done  for  Cepheid
variables,  for  example).   It   is also  necessary   to  correct for
obscuration due to dust (although in general, at least in the halos of
galaxies, these effects are likely to be small; a minor worry might be
that the  properties of  the dust could  evolve over time). In detail,
establishing   accurate K-corrections for    high-redshift supernovae,
measuring   reddenings,   and   correcting for  potential evolutionary
effects  will   be challenging,  although,     with the  exception  of
measurements   of   the  cosmic  microwave   background   anisotropies
(discussed in $\S$9 below), type Ia  supernovae may offer the best
potential for measuring $\rm \Omega_m$ and $\rm
\Omega_\Lambda$.

The most recent results based on  type Ia  supernovae (Perlmutter {\it
et al.} 1997 are encouraging, and they demonstrate that rapid progress
is likely  to be made in  the near  future.  Currently,  the published
sample size is limited to  7 objects;  however, many more objects have
now   been   discovered.   The   feasibility  of   discovering   these
high-redshift supernovae with high efficiency has  unquestionably been
demonstrated    ({\it e.g.} Perlmutter,    this    volume).   However,
systematic  errors are likely to  be  a significant component  of  the
error budget in the early stages of this program.

\section{ Summary of Current $\Omega_m$ and $\Omega_\Lambda$ Measurements}

The  results  of the preceding   sections on $\rm   \Omega_m$  and $\rm
\Omega_\Lambda$ are summarized  graphically in Figure 2. The 
diagonal dashed line denotes a flat ($\rm \Omega_m + \Omega_\Lambda$ =
1)  Universe.  Plotted  are  the results  from  dynamical measurements
(rotation curves,  Local Group dynamics,  galaxy velocity dispersions,
X-ray clusters) that tend to give low values of $\Omega
\sim$ 0.2-0.3.    In  addition, the  preliminary  results    from  the
Perlmutter {\it  et al.} (1997) type Ia  supernova  search are plotted
with quoted  1$\sigma$   error  bars, along  with  the     95\% limits
($\Omega_\Lambda <$ 0.66) on  $\rm \Omega_m$ and $\rm  \Omega_\Lambda$
from gravitational lens statistics from Kochanek (1996), shown as an
arrow along the diagonal.

\begin{figure} 
\psfig{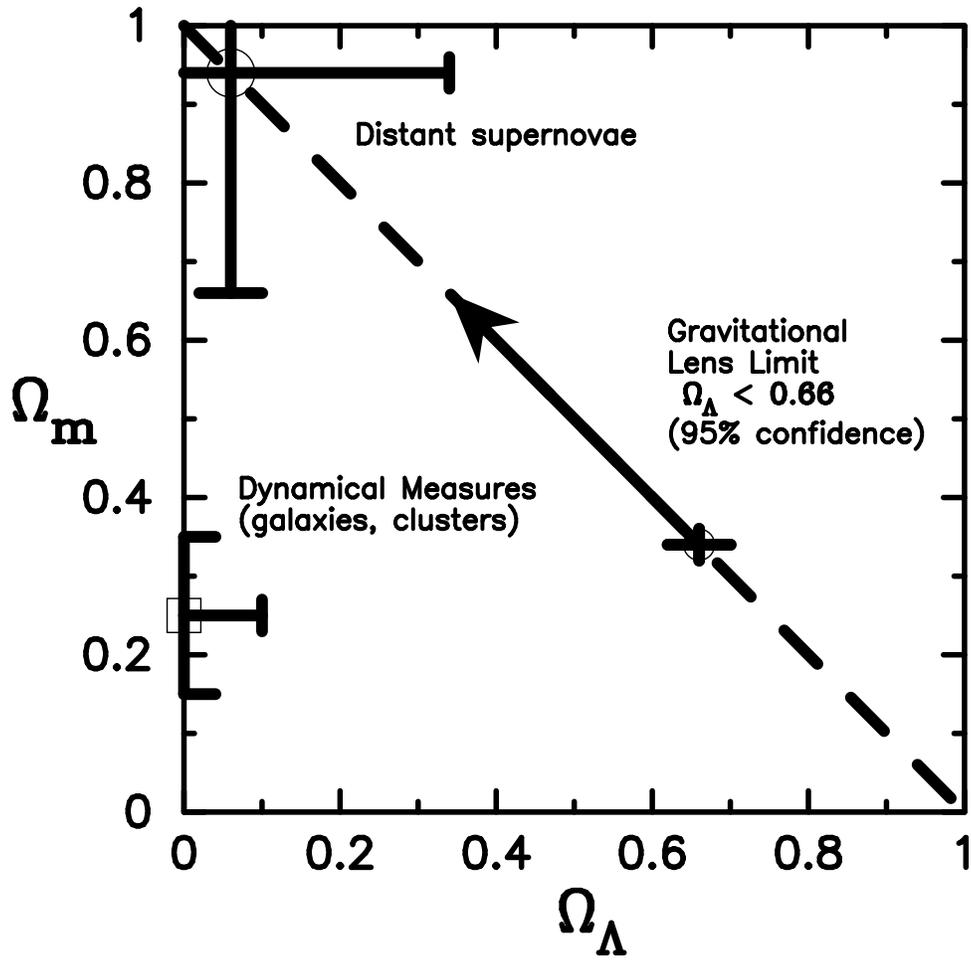} \caption{
Summary of Omega Determinations. The dashed line corresponds to
the case for a flat Universe: ($\rm \Omega_m + \Omega_\Lambda$ =
1). See text for details. }
\end{figure}

What can be concluded about the value of $\Omega$? Given the available
evidence and the  remaining uncertainties, plus underlying assumptions
at the present time, in my own view the data are still consistent with
both an open and a flat Universe.   This undesirable situation is very
likely to be resolved in the near future with more accurate mapping of
the anisotropies in the  cosmic  microwave  background radiation  (see
$\S$9).   At this  point in time,  however,  I believe that it is
premature either   to  sound  the   death  knell  for   (``standard'')
inflationary theories or to  conclude contrarily that an open Universe
is not a viable option.

\section{ $\bf H_0$ -- The Hubble Constant}

Sandage (1995) likens the measurement of H$_0$ to a game of chess.  In
chess,  only  a grand master `` experiences a compelling sense of the
issue and the   best move.  This  player ``knows''  by intuition which
clues are relevant...  In other words his or her intuition judges what
is real in the game, what will or will not lead to  contradiction, and
what aspects of the data to ignore.''

Although  there  are  perhaps differences   in  philosophy   and  many
different techniques for measuring $\rm H_0$, its importance cannot be
underestimated.  Knowledge of $\rm  H_0$ is  required to constrain the
estimates  of the baryon  density from nucleosynthesis at early epochs
in the Universe.  The larger the  value  of $\rm H_0$, the  larger the
component  of non-baryonic dark  matter is required, especially if the
Universe has a critical  density.  The Hubble  constant specifies both
the  time and length  scales at the  epoch  of equality of  the energy
densities of matter and radiation.  Both  the scale at the horizon and
the matter density determine the peak  in the perturbation spectrum of
the  early universe.   Hence,  an  accurate knowledge  of  the  Hubble
constant  can  provide powerful    constraints on   theories  of   the
large-scale structure  of galaxies.   At present, large values of $\rm
H_0$ are problematic for the  currently most successful models,  those
dominated by cold dark matter.

A value of $\rm H_0$ to $\pm$1\%  accuracy is still  a goal far beyond
currently available measurement techniques. However, if, for example a
value of $\rm H_0$ =  70 km/sec/Mpc  were confirmed  at $\pm$1\% (95\%
confidence), {\em and} the ages of the oldest objects in  the Universe
were  confirmed  to  be $>$12 Gyr, then  a number of issues   would be
brought into tight focus (and  corresponding new problems raised!).  A
cosmological constant would   be  {\em required}, there   would be  no
further debate over  the need  for  non-baryonic dark  matter,  and at
least the  standard  version of cold  dark  matter would  be ruled out
(conclusively).

The  requirements for  measuring  an accurate value of  $\rm  H_0$ are
simple to   list in  principle,  but extremely  difficult   to meet in
practice.  As discussed in more detail in Freedman (1997), in general,
there are 4 criteria that  need to be met  for any method.  First, the
method should be based upon well-understood physics; second, it should
operate well into  the smooth Hubble  flow (velocity-distances greater
than 10,000, and preferably, 20,000 km/sec); third, the  method should
be based on a statistically significant sample of objects, empirically
established to have  high internal accuracy;  and finally,  the method
needs to be demonstrated empirically  to be free of systematic errors.
This list of criteria applies both to classical distance indicators as
well as  to other physical methods (in  the  latter case, for example,
the  Sunyaev Zel'dovich  effect  or gravitational  lenses).   The last
point  requires that several   distance  indicators meeting the  first
three  criteria be  available,   but  the  current   reality is  that,
unfortunately,  at the present time, an   ideal  distance indicator or
other method  meeting all of the  above  criteria  does not exist. The
measurement of $\rm H_0$  to $\pm$1\% is  not yet  possible;  however,
recent  progress  (reviewed below)  illustrates that  a measurement to
$\pm$10\% is now feasible.

\subsection{ ``Physical'' versus ``Astronomical'' Methods}

There is  a common (mis)perception  that  some methods for determining
H$_0$ based on simple physical principles are  free  from the types of
systematics that often affect distance indicators (``physical'' versus
``astronomical'' methods).  However,  the fact remains that aside from
nearby   geometric parallax   measurements   (d$<$  100   pc),    {\it
astrophysics  enters all distance  and  H$_0$  determinations}!  These
methods include the gravitational lens time  delay method, the Sunyaev
Zel'dovich methods for clusters of galaxies, and theoretical modeling
of type Ia and II supernovae.

For   example,  it is certainly true  that   the gravitational lensing
method  is premised  on very  solid  physical principles  ({\it  e.g.}
Refsdael 1964,1966;  Blandford  \& Narayan 1992).   Unfortunately, the
astronomical  lenses  are   not idealized  systems  with  well-defined
properties that can  be  measured in a  laboratory; they are  galaxies
whose  underlying  (luminous  or dark)  mass   distributions are   not
independently  known, and furthermore   they may be  sitting in   more
complicated group or  cluster potentials.  A degeneracy exists between
the mass distribution of the lens and the value  of H$_0$ ({\it e.g.,}
Kundi\'{c} {\it et al.} 1997; Keeton and Kochanek 1997;  Schechter {\it et
al.} 1997).  This is not a method  based  solely on well-known physics;
it is a method that also requires knowledge  of astrophysics.  Ideally
velocity dispersion measurements as a function  of position are needed
(to constrain  the mass distribution of  the lens).  Such measurements
are  very difficult (and generally  have not been available).  Perhaps
worse yet, the distribution  of  the  dark matter in these  systems is
unknown.  In a similar way, the Sunyaev-Zel'dovich method is sensitive
to the clumping of X-ray  gas, discrete radio sources, the  projection
of the clusters, and other astrophysical complications.

Hence the methods for measuring H$_0$ cannot be cleanly separated into
purely   ``physical''  and ``astronomical''  techniques.  Rather, each
method has its own set of advantages and disadvantages. In my view, it
is  vital to  measure  H$_0$ using a variety  of  different methods in
order to  identify potential systematic  errors  in any one technique.
All methods require large, statistically significant samples.  This is
one of  the current  weakest  aspects of   the Sunyaev-Zel-dovich  and
gravitational-lens methods,  for  example, where samples  of  a few or
only 2  objects, respectively, are  currently available.  In contrast,
it   is  a   clear disadvantage  that  many of  the classical distance
indicators ({\it e.g.}, the Tully-Fisher relation and at present, even
the type Ia supernovae) do not have a  well-understood physical basis.
However,  there are    many cross-checks  and    tests  for  potential
systematic effects that are now feasible and are being carried out for
large  samples of    measured  extragalactic  distances (see  $\S$5.4
below).  Assuming that systematic effects can eventually be understood
and  minimized,  ultimately,  the   measurement  of $\rm   H_0$   by a
geometrical (or optical) technique at large distances will  be crucial
for establishing the reliability of the classical distance scale.  For
gravitational lenses,  however, a considerable  amount of work will be
required to increase the numbers of systems with measured time delays,
obtain velocity  dispersion profiles for  the faint lensing  galaxies,
constrain  the lens  models and test  for other systematic effects, if
this goal is to be reached.

Below,  progress on $\rm  H_0$   measurements based  on  gravitational
lenses, the Sunyaev Zel'dovich effect, and the  extragalactic distance
scale is briefly summarized.

\subsection{ Gravitational Lenses}

Refsdael (1964, 1966) noted that the  arrival times for the light from
two gravitationally lensed images of   a background point source   are
dependent    on  the path   lengths and   the  gravitational potential
traversed in each case. Hence, a measurement of the time delay and the
angular  separation for different  images of a  variable quasar can be
used to provide    a measurement of  $\rm H_0$.     This method offers
tremendous potential because it can be applied  at great distances and
it is  based on very solid  physical principles.  Moveover, the method
is  not very sensitive  to  $\rm \Omega_m$  and  $\rm \Omega_\Lambda$.
Some of the  practical  difficulties   in  applying this  method  have
already been discussed in the previous section.

A number   of  new results  based   on this technique   have  recently
appeared.  Estimates of time delay measurements are  now available for
2 systems: 0957 +561 (Kundi\'{c} {\it et  al.} 1997), and most recently, a
new time delay has been measured  for PG 1115  (Schechter {\it et al.}
1997; Keeton and Kochanek 1997).

In  the  case of  0957+561, progress has  been made on several fronts.
The time delay for this system has been a matter of some debate in the
literature, with  two different   values  of 410 and   536  days being
advocated; extensive new optical data have now resolved this issue in
favor of the  smaller  time delay  ($\Delta  t$=417$\pm$3 days (Kundi\'{c}
{\it et al.} 1997).  Another large observational  uncertainty has been
due to the difficulty of measuring an accurate velocity dispersion for
the lensing galaxy.  Recent data from the Keck telescope have provided
a  new  measurement of  the  velocity dispersion  (Falco {\it et  al.}
1997).  In addition, there has been substantial  progress in modeling
this system (Grogin \& Narayan 1996).  Based on the new time delay and
velocity dispersions measurements, and the model of Grogin and Narayan,
Falco {\it et al.} have recently derived a value of $\rm H_0$ = in the
range  62 - 67 $\pm$  8  km/sec/Mpc  for   this system.   The velocity
dispersion in the lensing galaxy appears to decrease very steeply as a
function  of  position  from  the   center  of  the  galaxy;   further
higher-resolution   measurements will be   required  to  determine the
reliability of these faint measurements.

Schechter {\it   et  al.} 1997 have  undertaken  an  extensive optical
monitoring  program  to measure two  independent  time delays  in  the
quadruply-imaged quasar PG 1115+080.  They fit a  variety of models to
this system, preferring a solution that yields a value of  $\rm H_0$ =
42 km/sec/Mpc $\pm$  14\% (for $\rm  \Omega$ = 1).  The  model in this
case consists of fitting isothermal spheres to both the lensing galaxy
and a nearby group of galaxies. They also considered additional models
that yield values of $\rm  H_0$  = 64  and 84   km/sec/Mpc. Keeton  \&
Kochanek (1997) have considered a  wider class  of models. They stress
the degeneracies that are inherent in  these analyses; a number of
models   with differing radial  profiles  for  the lensing galaxy  and
group, and with differing  positions  for  the group,  yield fits with
chi-squared per degrees  of freedom less  than  1. They conclude  that
$\rm H_0$ = 60 $\pm$ 17 km/sec/Mpc (1-$\sigma$).

\subsection{ Sunyaev Zel'dovich Effect and X-Ray Measurements}

The  inverse-Compton scattering of photons from   the cosmic microwave
background off of hot  electrons in the  X-ray  gas  of  rich clusters
results in a measurable decrement in the microwave background spectrum
known as  the Sunyaev-Zel'dovich  (SZ) effect (Zel'dovich and Sunyaev  
1969).  Given a spatial (preferably 2-dimensional) distribution of the
SZ effect and a high-resolution X-ray map, the density and temperature
distributions  of the hot  gas  can  be  obtained;  the mean  electron
temperature can  be obtained from  an X-ray  spectrum.  An estimate of
H$_0$ can be made based on the definitions of the angular-diameter and
luminosity distances.  The method makes use of the fact that the X-ray
flux is distance-dependent,  whereas the Sunyaev-Zel'dovich  decrement
in the temperature is not. 


Once again, the advantages of this method  are that it can  be applied
at large  distances   and,  in principle, it  has   a  straightforward
physical  basis.   As   discussed  in   $\S$5.1, some  of  the    main
uncertainties with this method are due to potential  clumpiness of the
gas (which would result in reducing $\rm H_0$), projection effects (if
the clusters observed  are prolate, $\rm  H_0$ could  be larger),  the
assumption of hydrostatic  equilibrium, details of  the models for the
gas and   electron  densities, and  potential contamination from point
sources.

To date, a range of values of $\rm  H_0$ have  been published based on
this method ranging from $\sim$25 - 80 km/sec/Mpc ({\it e.g.}, McHardy
{\it  et al.} 1990; Birkinshaw \&  Hughes 1994; Rephaeli 1995; Herbig,
Lawrence \& Readhead 1995). The uncertainties are still  large, but as
more and  more clusters are observed,  higher-resolution (2D)  maps of
the  decrement,   and  X-ray maps  and spectra  become  available, the
prospects  for   this  method  will   continue  to  improve.  At  this
conference, Carlstrom reported  on a  new  extensive  survey of lenses
being  undertaken  both  at   Hat Creek  and  the Owens   Valley Radio
Observatory.  X-ray  images are being  obtained with   ROSAT and X-ray
spectra with ASCA.

\subsection{ The Cepheid-Calibrated Extragalactic Distance Scale}

Establishing accurate extragalactic distances has  provided an immense
challenge to astronomers since the 1920's. The situation has improved
dramatically as better  (linear) detectors have become available,  and
as several  new, promising techniques have  been  developed.   For the
first time in the history of this  difficult field, relative distances
to  galaxies  are being compared on  a  case-by-case  basis, and their
quantitative  agreement is    being  established.  Several,   detailed
reviews  on  this progress have  been  written  (see, for example,  the
conference  proceedings  for the  Space Telescope   Science  Institute
meeting on  the  Extragalactic Distance   Scale edited by  Donahue and
Livio 1997).

The  Hubble  Space Telescope (HST)   Key Project  on   H$_0$  has been
designed  to undertake  the  calibration of   a number  of   secondary
distance methods using Cepheid variables (Freedman  {\it et al.} 1994;
Kennicutt, Freedman \& Mould 1995; Mould {\it et al.} 1995).  Briefly,
there are  three primary goals: (1) To  discover Cepheids, and thereby
measure  accurate  distances  to   spiral  galaxies suitable  for  the
calibration of several   independent secondary methods.   (2)  To make
direct  Cepheid measurements of distances  to three spiral galaxies in
each  of the  Virgo and Fornax  clusters.   (3)  To provide a check on
potential systematic errors both in the Cepheid distance scale and the
secondary methods. The final  goal is to derive   a value for the  the
Hubble constant, to  an  accuracy of  10$\%$.  Cepheids are also being
employed  in several other  HST distance scale  programs  ({\it e.g.},
Sandage  {\it et al.}  1996; Saha {\it  et al.} 1994,  1995, 1996; and
Tanvir {\it et al.} 1995).

In Freedman,   Madore  \&  Kennicutt (1997),  a  comparison of Cepheid
distances   is made  with   a    number of  other   methods  including
surface-brightness  fluctuations,  the  planetary nebula    luminosity
function, tip  of   the  red giant   branch, and  type II  supernovae.
(Extensive  recent  reviews  of all of these  methods can be  found in
Livio and Donahue (1997); by Tonry; Jacoby; Madore, Freedman \& Sakai;
Kirshner).  In general,  there is   excellent agreement amongst  these
methods; the relative distances  agree to within  $\pm$10\% (1-sigma).
The use of both type Ia and type  II  supernovae for  the  purposes of
determining $\rm H_0$ are described in this volume by Filippenko.

The results of the $\rm H_0$ Key Project have been summarized recently
by Freedman, Madore  \& Kennicutt  (1997); Mould {\it et  al.} (1997);
and Freedman  (1997). For somewhat   different views, see  Sandage  \&
Tammann (1997).  The  remarks  in  the rest of  this    section follow
Freedman (1997). At this mid-term  point  in the  HST Key Project, our
results  yield a value  of H$_0$  = 73 $\pm$ 6  (statistical) $\pm$  8
(systematic) km/sec/Mpc. This result is based on  a variety of methods,
including a Cepheid calibration of the Tully-Fisher relation,  type Ia
supernovae, a  calibration  of  distant clusters  tied  to Fornax, and
direct Cepheid distances out to $\sim$  20~Mpc.  In Table 2 the values
of H$_0$ based on these various methods are summarized.

\medskip

\begin{table} 
  \begin{center} 
  \caption{SUMMARY OF KEY PROJECT RESULTS ON H$_0$}
\smallskip
  \begin{tabular}{lc} 
             &         \\
      Method & H$_0$   \\[3pt] 
       Virgo & 80 $\pm$ 17 \\ 
       Coma via Virgo   & 77 $\pm$ 16 \\
       Fornax & 72 $\pm$ 18 \\ 
       Local  & 75 $\pm$ ~8 \\ 
       JT clusters & 72 $\pm$ ~8 \\
       SNIa   & 67 $\pm$ ~8 \\ 
       TF     & 73 $\pm$ ~7 \\ 
       SNII   & 73 $\pm$ ~7 \\ 
       D$_N -\sigma$ & 73 $\pm$ ~6 \\ 
              &   \\
       Mean   &   73 $\pm$ ~4 \\
              &   \\
     
       {\bf Systematic Errors} & {\bf ~~~$\pm$ 4 ~~~~~$\pm$ 4 ~~~~~$\pm$ 5  ~~~~~$\pm$ 2} \\
              & ~~~(LMC) ~([Fe/H]) ~(global) ~(photometric) \\
              &   \\
  \end{tabular}

  \caption{Current values of   H$_0$  for various   methods.    For each
method, the formal statistical uncertainties are given. The systematic
errors (common to all of these  Cepheid-based calibrations) are listed
at the  end of  the table.   The dominant uncertainties  are   in  the
distance to  the LMC and  the potential effect  of  metallicity on the
Cepheid period-luminosity relations, plus an allowance is made for the
possibility that the locally  measured value of H$_0$  may  differ from
the global value.  Also allowance is made for a systematic scale error
in the photometry which  might be affecting  all software packages now
commonly in use.  Our best current  weighted mean value  is H$_0$ = 73
$\pm$ 6 (statistical) $\pm$ 8 (systematic) km/sec/Mpc.}
\end{center}
\end{table}

\medskip

These  recent results on  the  extragalactic distance scale  are  very
encouraging.  A  large   number   of  independent   secondary  methods
(including the most recent  type  Ia supernova calibration by  Sandage
{\it et al.} 1996) appear to be converging on a value of H$_0$  in the
range   of 60  to    80 km/sec/Mpc.  The   long-standing factor-of-two
discrepancy in H$_0$ appears to be  behind us.  However, these results
underscore the importance of reducing remaining  errors in the Cepheid
distances  ({\it e.g.,}   those  due  to reddening   and   metallicity
corrections), since at present the majority of distance estimators are
tied in zero point to the Cepheid distance  scale.  A 1-$\sigma$ error
of $\pm$10\% on H$_0$ (the aim of the Key  Project)  currently amounts
to  approximately $\pm$ 7    km/sec/Mpc, and translates into   a  95\%
confidence interval on $\rm H_0$ of roughly 55 to 85 km/sec/Mpc.

While   this   is  an   enormous  improvement   over the factor-of-two
disagreement of the previous decades, it  is not sufficiently precise,
for  example, to discriminate  between current models   of large scale
structure formation,  to  resolve definitively    the  fundamental age
problem, or to settle  the question of a  non-zero value of $\Lambda$.
Before compelling constraints can  be made on  cosmological models, it
is  imperative  to rule out  remaining sources of systematic error  in
order to severely limit the  alternative  interpretations  that can be
made of the data.  The spectacular success of HST, and the fact that a
value  of H$_0$   accurate  to 10\%   (1-$\sigma$)   now appears quite
feasible, also brings  into  sharper  focus smaller (10-15\%)  effects
which  were buried in  the    noise during  the  era  of factor-of-two
discrepancies.    Fortunately,  a  significant   improvement  will  be
possible  with the new infrared  capability  afforded by  the recently
augmented near-infrared  capabilities of HST (the NICMOS  instrument).
Planned NICMOS  observations will reduce   the remaining uncertainties
due to both reddening and metallicity by a factor of 3.

\section{ $\bf t_0$  - Ages of the Oldest Stars}

The   ages  of stars can    be  derived quite  independently from  the
expansion age of the  Universe (obtained by  integrating the Friedmann
equation), and  have  long been  used as  a   point of comparison  and
constraint on  cosmology;  for example,  globular cluster  age-dating,
nucleocosmochronology,  and  white-dwarf  cooling   estimates for  the
Galactic  disk.   The reader  is referred to earlier  reviews on these
topics by  Renzini (1991), Schramm   (1989). For the purposes of  this
review, I briefly consider only two types of age determinations: those
based on Galactic  globular clusters,  and a new estimate  of the  age
based on a measurement of radioactive thorium in a metal poor Galactic
halo star.

\subsection{ Globular Cluster Ages }

There are also many excellent recent reviews  covering in great detail
the ages obtained for  Galactic globular clusters  ({\it i.e.,} from a
comparison  of   observed  color magnitude  diagrams   and theoretical
evolution models).  At the moment, there is  a fairly  broad consensus
that Galactic globular clusters are most likely at least 14-15 Gyr old
({\it e.g.} Chaboyer {\it et al.} 1996; VandenBerg {\it  et al.} 1996;
Shi 1995).

It is not widely appreciated that {\it the largest uncertainty  in the
globular-cluster ages results  from uncertainties in the  distances to
the  globular clusters},   which  currently are  based on  statistical
parallax measurements of Galactic RR Lyrae  stars or on parallaxes for
nearby subdwarfs({\it e.g.} Renzini, 1991; Chaboyer {\it et al.} 1996;
VandenBerg {\it et al.} 1996).  Although the ages of globular clusters
are widely regarded  as theoretically-determined  quantities,  in  the
process of determining ages, it is still necessary to interface theory
with  observation  and   transform   the   observed globular   cluster
magnitudes  to   bolometric  luminosities  (via an   accurate distance
scale).   The  subdwarf  and RR   Lyrae statistical parallax  distance
calibrations   currently differ     by   about  $\sim$0.25-0.30   mag.
Unfortunately, as  emphasized  by  Renzini, small  errors in  distance
modulus (0.25 mag or 13\% in distance)  correspond to 25\% differences
in age.  Even  with improved parallax  measurements (for example, soon
to be available  from HIPPARCHOS), there are  many subtle issues ({\it
e.g.,} reddening, metallicity, photometric zeropoints) that combine to
make it a  very difficult problem  to achieve distances to better than
5\% accuracy.

As discussed previously  in many  contexts  ({\it  e.g.} Walker  1992;
Freedman  \& Madore  1993; van den Bergh  1995, and most  recently by
Feast \& Catchpole 1997), there is also currently a discrepancy in the
Cepheid and  RR Lyrae distances to nearby  galaxies.   If  the Cepheid
distances are correct, it would imply that the absolute  magnitudes of
RR   Lyraes are brighter   (by   about  0.3  mag)  than  suggested  by
statistical parallax and Baade-Wesselink  calibrations for Galactic RR
Lyraes ({\it e.g.} see VandenBerg, Bolte \& Stetson 1996  for a recent
discussion).  This  brighter RR Lyrae calibration  agrees well in zero
point  with that from  Galactic   subdwarfs.  Based  on  the models of
VandenBerg  {\it et   al.} 1997, applying  this  calibration (adopting
M$_V$(RR)=0.40 mag) to the metal-poor globular cluster M92, results in
an age of 15.8$\pm$2 Gyr.  If the  fainter  RR Lyrae distance scale is
correct, the  age derived for  M92 based on  these same  recent models
increases to  $\sim$19 Gyr.  Alternatively, if the Feast  \& Catchpole
calibration  of Galactic  Cepheids based on HIPPARCHOS  parallaxes is
correct,  then  the resulting RR Lyrae   calibration  is even brighter
(M$_V$(RR)=0.25 at [Fe/H] = -1.9),  and  the corresponding age for M92
would be   reduced to about  13 Gyr  (based on  the   same  Vandenberg
models).   A new  calibration of Galactic   metal-poor subdwarfs, also
based on new  HIPPARCHOS parallaxes, appears  to confirm these younger
ages  (Reid, private communication).   It is interesting  to  note that
while the distances to nearby galaxies have converged to a level where
they  no longer have  a  factor-of-two impact on  the Hubble constant,
subtle  differences of only a few  tenths of  a  magnitude in distance
modulus can  still have very  significant impact on cosmology, through
the ages determined from stellar evolution.

\subsection{ Thorium Ages }

A new measurement of the age of a very metal poor star in  the halo of
our  Galaxy has  recently been made  by  Cowan  {\it  et  al.} (1997),
following  a technique introduced by   Butcher (1987).  These authors
make  use of very  high-resolution echelle spectra   of CS22892-052, a
star with a metallicity of  only [Fe/H]  = $-$3.1. They find  that the
observed abundances  for  stable  elements  in this  star   match  the
observed r-process elemental abundances observed in the Sun.  However,
for the radioactive element thorium, the abundance is down by a factor
of  40  relative to  solar.    Allowing for the radioactive   decay of
thorium relative to  (stable) europium  yields a minimum age  for this
star of 15.2 $\pm$ 3.7 Gyr (1-sigma). If instead of europium alone, an
average abundance  for all r-process elements  from Eu-Er is  used, an
age of 13.8 $\pm$  3.7 Gyr results.   This lower limit to  the  age is
independent  of any model of  Galactic evolution (which only  serve to
increase the  total  age estimates for the  Universe).   It depends on
both  the decay  rate and the initial abundance  of thorium.  Although
the current   sample is small (1  star!)    and the  uncertainties are
correspondingly large, there is  excellent promise for the future once
the  sample  is enlarged.   Methods  like  this  one  are particularly
important because of  the  opportunity   of having high-quality   ages
completely independent of the globular cluster age scale.

\section{ Remaining Issues for Measuring $\bf t_0$}

What  are the  ages of the oldest objects  in the Universe?   In  this
context, we  need to keep in mind  that it is currently  only a useful
working   hypothesis   that  the    Galactic globular    clusters  are
representative  of the oldest objects in  the Universe ({\it e.g.} see
Freedman 1995 for a more detailed  discussion).  Currently, the sample
of objects  for which  direct ({\it  i.e.} main-sequence-fitting) ages
can  be measured is  limited to our  own Galaxy  and a small number of
satellites around our own Galaxy.  It is  at least conceivable that in
denser environments in the early  Universe, star formation  could have
proceeded earlier than for Galactic globular  clusters.  At this time,
there is   no direct information  with   which to constrain  the  true
dispersion in  (or upper  limit  to) ages in environments  outside the
nearest galaxies in our  own Local  Group.  There are, for example, no
giant elliptical galaxies  in the Local  Group.  Although considerable
effort is now  being invested in finding  potential  ways to lower the
Galactic globular cluster ages, there is  reason to keep  in mind that
the expansion-age discrepancy could potentially be  even worse than is
currently being discussed.

\section{ $\bf H_0t_0$}

One of the most powerful tests for a non-zero cosmological constant is
provided by a  comparison of  the expansion and  oldest-star ages.  To
quote Carroll, Press and Turner (1990),  ``A high value of H$_0$ ($>$80
km/s/Mpc, say), combined with  no loss  of confidence in a value 12-14
Gyr  as   a   {\it minimum} age   for  some   globular clusters, would
effectively prove the  existence of  a significant $\rm \Omega_\Lambda$
term.   Given  such observational results,  we  know  of no convincing
alternative hypotheses."


In Figure 3, the dimensionless product of $\rm H_0t_0$ is plotted as a
function of $\rm \Omega$.  Two different  cases  are illustrated:
an open $\rm \Omega_\Lambda$ = 0 Universe, and  a flat Universe with 
$\rm \Omega_\Lambda + \Omega_m$ = 1.  Suppose that both $\rm  H_0$ and
$t_0$ are    both  known to  $\pm$10\%    (1-$\sigma$, {\it  including
systematic   errors}).   The  dashed  and   dot-dashed  lines indicate
1-$\sigma$ and 2-$\sigma$  limits, respectively for  values of H$_0$ =
70 km/sec/Mpc and t$_0$ = 15 Gyr.  Since the two  quantities H$_0$ and
t$_0$ are completely independent, the  two errors have  been added  in
quadrature,  yielding a  total  uncertainty   on the product  of  $\rm
H_0t_0$ of $\pm$14\%  $rms$.  These values  of $\rm H_0$ and $\rm t_0$
are  consistent   with a Universe   where $\rm  \Omega_\Lambda$ = 0.8,
$\Omega_m$ =  0.2.   The  Einstein-de  Sitter model   ($\Omega_m$  =1,
$\Omega_\Lambda$=0) is excluded (at 2.5$\sigma$).

Despite  the enormous progress  recently in the   measurements of $\rm
H_0$  and $\rm t_0$, Figure  3  demonstrates  that significant further
improvements are still needed.  First, in the opinion of  this author,
{\it total} (including both statistical  and systematic) uncertainties
of  $\pm$10\% have yet to be  achieved for either   $\rm  H_0$ or $\rm
t_0$.  Second, assuming  that such accuracies  will be forthcoming  in
the near future for $\rm H_0$  (as the Key Project, supernova programs
and other  surveys near   completion),  and for $t_0$  (as  HIPPARCHOS
provides an improved calibration both for RR Lyraes and subdwarfs), it
is clear from this   figure  that if  $\rm H_0$   is as high   as   70
km/sec/Mpc, then accuracies  of significantly  {\it better than} $\pm$
10\% will be    required to rule in   or  out  a non-zero   value  for
$\Lambda$.  (If $\rm H_0$ were larger (or  smaller), this discrimination
would be simplified!)

\begin{figure} 
\psfig{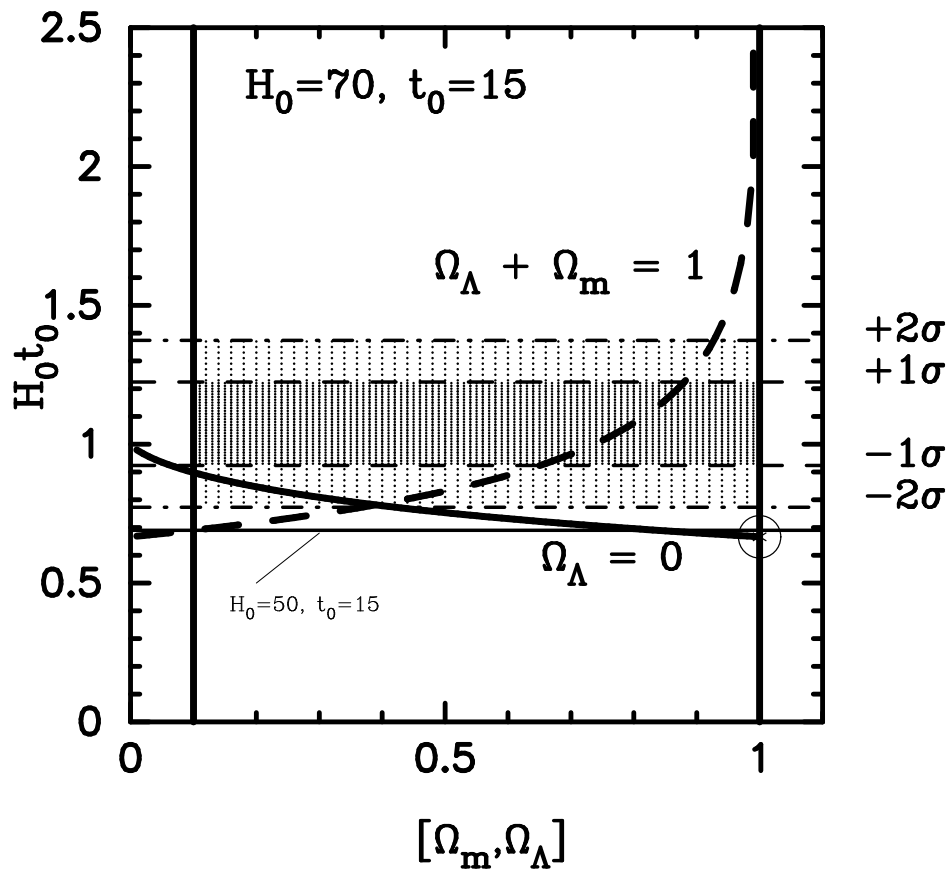} 
\caption{ The product of $\rm H_0t_0$ as a  function of
$\rm \Omega$.  The dashed curve indicates the case of  a flat Universe
with $\rm \Omega_\Lambda + \Omega_m$  = 1.  The  abscissa in this case
corresponds  to  $\rm \Omega_\Lambda$.  The   solid curve represents a
Universe with $\rm  \Omega_\Lambda$ = 0.     In  this  case, the abcissa
should  be  read as $\rm   \Omega_m$. The dashed  and dot-dashed lines
indicate 1-$\sigma$ and 2-$\sigma$ limits,  respectively for values of
H$_0$ =  70  km/sec/Mpc  and  t$_0$ =  15  Gyr in the case  where both
quantities are known to $\pm$10\% (1-$\sigma$).  The large open circle
denotes values  of  H$_0$t$_0$ = 2/3 and  $\Omega_m$  = 1 ({\it i.e.,}
those predicted by the standard Einstein-de  Sitter model). Also shown
for comparison is a  solid line for  the  case H$_0$ = 50  km/sec/Mpc,
t$_0$ = 15 Gyr.
\label{fig:radish}}
\end{figure}



\section{ Cosmological Parameters from Cosmic Microwave Background
Anisotropies}

One  of  the most  exciting  future developments with  respect  to the
accurate   measurement  of  cosmological   parameters   will   be  the
opportunity to measure anisotropies in the cosmic microwave background
to high precision.  Planned balloon-born experiments ({\it e.g.}, MAX,
MAXIMA, and Boomerang) will shortly  measure the position of the first
acoustic peak in the cosmic background anisotropy spectrum.  Even more
promising are   future satellite  experiments ({\it  e.g.},  MAP  to be
launched by NASA in  2000, and the  European COBRAS/SAMBA mission, now
renamed the PLANCK Surveyor mission, currently planned to be launched in 2005).

The underlying physics governing the shape of  the anisotropy spectrum
is that describing the interaction  of  a very  tightly coupled  fluid
composed of electrons and photons before  (re)combination ({\it e.g.},
Hu  \& White 1996; Sunyaev \&  Zel'dovich 1970).   It is elegant, very
simple in principle, and  offers  extraordinary promise for  measuring
cosmological parameters;  ({\it e.g.}, $\rm H_0$, $\rm  \Omega_0$, and
the  baryon  density  $\rm \Omega_b$  to precisions  of 1\% or better:
Bond, Efstathiou  \&  Tegmark 1997).   

The final accuracies will of course (again) depend on how well various
systematic  errors  can   be controlled  or     eliminated.  The major
uncertainties will be determined by how well foreground sources can be
subtracted,  and probably  to a   lesser  extent, by  calibration  and
instrumental uncertainties.  (PLANCK will provide a cross check of the
MAP calibration.)  Potentially the greatest  problem is the  fact that
extracting cosmological parameters  requires a specific model  for the
fluctuation spectrum.  Currently the estimates of the precisions ({\it
i.e.,} without  systematic effects included)   are based on models  in
which the primordial fluctuations are Gaussian  and adiabatic, and for
which there is no preferred scale.  A  very different anisotropy power
spectrum shape  is predicted for   defect  theories (Turok  1996), but
these calculations are  more difficult and have not   yet reached  the
same level of predictive power.  Important additional constraints will
come from  polarization measurements {{\it e.g.}, Zaldarriaga, Spergel
\& Seljak 1997; Kamionkowski {\it et al.} 1997). The polarization data
will provide a means of breaking  some of the degeneracies amongst the
cosmological  parameters  that   are  present in  the temperature data
alone. Furthermore, they  are  sensitive to  the presence of a  tensor
(gravity wave) contribution,  and hence will   allow a very  sensitive
test of inflationary models.

Figure 4 shows a  plot of  the  predicted angular power  spectrum  for
cosmic microwave  background  (CMB) anisotropies  reproduced from  Hu,
Sugiyama, \& Silk (1997).  The position of the first  acoustic peak is
very  sensitive to the value of  $\Omega_0$, and,   as  noted by these
authors, the spacing between the acoustic  peaks in the power spectrum
appears  to provide   a fairly robust   measure  of  $\Omega_0$.   The
accurate determination of  other cosmological parameters  will require
the measurement of peaks  at  smaller (arcminute)  angular scales.  In
general, the ratio of the first to the third peaks is sensitive to the
value of value of $H_0$ ({\it e.g.,}  Hu \& White 1996). Excellent sky
coverage is critical to these efforts in  order to reduce the sampling
variance.

\begin{figure} 
\plotfiddle{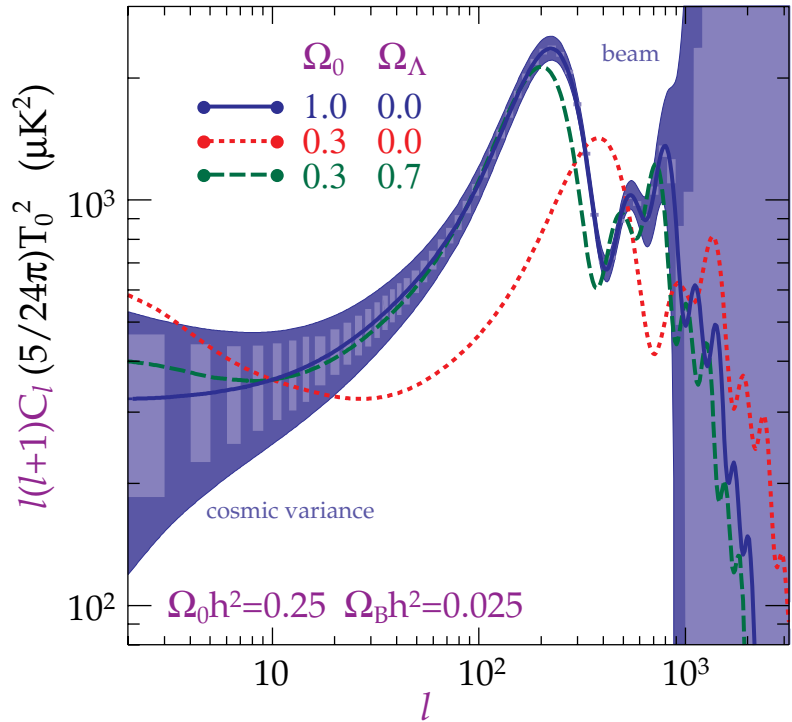}{5.in}{270}{50}{50}{-210}{350}
\caption{  The    angular   power spectrum  of    cosmic
microwave      background anisotropies   assuming  adiabatic,   nearly
scale-invariant  models for a  range  of values of $\rm  \Omega_0$ and
$\rm  \Omega_\Lambda$ (Hu, Sugiyama,  and  Silk 1997; their Figure  4).
The C$_l$ values correspond to
the squares of the spherical harmonics  coefficients.   Low $l$ values
correspond to large  angular scales ($l  \sim { 200\deg \over \theta }
$). The position of the first acoustic peak is predicted  to  be at $l
\sim $220$\Omega_{TOT}^{-1/2}$,  and hence, shifts to  smaller angular
scales for open universes.  }
\end{figure}

Can the  cosmological parameters  be measured  to precisions  of $\leq
1\%$  with currently   planned experiments  as   advertised  above?  I
believe that both MAP and  PLANCK	 are likely to  revolutionize
our understanding of  cosmology.  Observation of a Gaussian, adiabatic
fluctuation  spectrum  would  be   a   stunning confirmation   of  the
``standard''  cosmology.   However, equally   fundamental would be the
case where the  observed  anisotropy spectrum resembles   nothing like
those for any of the various current  theoretical predictions.  In the
former case,  {\it if} foreground effects can  be accounted  for, then
measurement   of the cosmological   parameters  to  these   levels  of
precision will eventually follow.   However,  in  the latter case,  at
least until the origin of  the spectrum could be predicted  from first
principles,  all   bets  would  be   off  for   the determination   of
cosmological parameters.

Can the foreground subtraction be  accounted for accurately enough  to
yield final accuracies of 1\%  (or better)?  There will  be foreground
contributions due to faint,  diffuse Galactic emission.  MAP will have
5 frequency bands ranging from 22 to 90 GHz allowing both the spectral
and  spatial distribution of  the Galactic  foreground to be measured.
PLANCK will  have 9  frequency channels    from  30 GHz  to  900  GHz.
However,  there are many sources of   foregrounds whose subtraction is
critical; perhaps the  greatest  unknown is the potential contribution
from  GHz radio sources, many   of  which could  potentially  also  be
variable sources.  Deep  90 GHz  radio surveys  from the ground  might
address  the question  of how serious an  issue  such sources could be
(Spergel, private communication). Although  MAP will  cover  any given
region of the sky several times, the signal-to-noise for an individual
image will be  insufficient  to detect any  but the brightest sources.
In addition there   will be foreground contributions  due   to diffuse
emission  from external galaxies,   dust within  galaxies,  and bright
infrared luminous galaxies.  Until these experiments are completed, it
will be difficult to assess whether these systematic uncertainties are
likely to be small relative to the quoted formal uncertainties.

\section{ Summary }

The current best measurements for the cosmological parameters yield:

\smallskip

\hskip 0.5cm $\rm \Omega_m \sim$ (0.2 - 0.4) $\pm$ 0.1 ~~~~~~~~~~~~~~~(1-$\sigma$)

\hskip 0.5cm $\rm H_0 \sim$ (67 - 73) $\pm$ 7 km/sec/Mpc  ~~(1-$\sigma$)

\hskip 0.5cm $\rm t_0 \sim$ (14 - 15) $\pm$ 2 Gyr  ~~~~~~~~~~~~~~~(1-$\sigma$)

\hskip 0.5cm $\rm \Omega_\Lambda <$ 0.7  ~~~~~~~~~~~~~~~~~~~~~~~~~~~~~~~~~(2-$\sigma$)
\smallskip

\noindent
The low value  for $\rm \Omega_m$ and  relatively high value  for $\rm
H_0t_0$ do not favor the standard Einstein-de Sitter ($\rm \Omega_m$ =
1, $\rm \Omega_\Lambda$ =  0)  Universe; however, this model cannot be
ruled out  at  high   statistical significance.  Moreover,  systematic
errors are still  a source of serious  concern.  If the new HIPPARCHOS
calibrations are confirmed,  the  ages of  globular clusters may be as
low as  10-12 Gyr.   Rapid  progress is expected in  addressing  these
systematic effects; in  particular new data  from HST, HIPPARCHOS, and
MAP/PLANCK offer the enticing possibility that all of the cosmological
parameters  may  soon  be  measured  to   unprecedented accuracies  of
$\pm$1-5\%  within a decade.   Let us  hope that unexpected systematic
errors will  not continue to lurk (as  they  have done historically so
many  times before)  in  these future  efforts  to  define  the  basic
cosmological parameters.

\section*{Acknowledgments} 
It is a  pleasure to thank  the organizing committee  for an extremely
enjoyable   and  interesting conference, and   for the opportunity  to
speak.  The work presented on the Cepheid-based extragalactic distance
scale ($\S$5.4) has been done in collaboration with the Hubble Space
Telescope Key Project team on  the Extragalactic Distance  Scale and I
would  like to acknowledge the contributions   of  R.  Kennicutt, J.R.
Mould  (co-PI's), S.  Faber,  L.  Ferrarese, H.   Ford, B.  Gibson, J.
Graham,  J.   Gunn, M.  Han, J.   Hoessel, J.   Huchra, S.  Hughes, G.
Illingworth, B.F.   Madore,  R.     Phelps, A.  Saha,  S.    Sakai, N.
Silbermann,  and P.  Stetson, and  graduate  students F.  Bresolin, P.
Harding, D.  Kelson, L.  Macri, D.  Rawson, and A.  Turner.  This work
is based on  observations  with the NASA/ESA  Hubble Space  Telescope,
obtained by the Space  Telescope Science Institute, which is  operated
by AURA,  Inc.  under  NASA contract  No.  5-26555.   Support for this
work was provided by NASA through grant GO-2227-87A from STScI.


\def\aj{\em Astron. J.}
\def\araa{\em Astron. Rev. Astron. Astrophys.}
\def\apj{\em Astrophys. J.}
\def\apjl{\em Astrophys. J. Lett.}
\def\apjs{\em Astrophys. J. Suppl.}
\def\apss{\em Astrophys. \& SS}
\def\aa{\em Astron. Astrophys.}
\def\aas{\em Astron. Astrophys. Suppl.}
\def\mnras{\em Mon. Not. Royal Astr. Soc.}
\def\nat{\em Nature,}
\def\pasp{\em Publ. Astr. Soc. Pac.}
\def\rmp{\em Rev. Mod. Phys.}
\def\sci{\em Science,}

\end{document}